\documentclass[aps,prb,twocolumn,superscriptaddress,showpacs]{revtex4}
\usepackage{graphicx}
\begin{document}
\title{\bf Quantum phase transitions in a new exactly solvable quantum spin
biaxial model with multiple spin interactions}

\author{A.A.~Zvyagin}
\affiliation{Max-Planck-Institut f\"ur Physik komplexer Systeme,
N\"othnitzer Str., 38, D-01187, Dresden, Germany}
\affiliation{B.I.~Verkin Institute for Low Temperature Physics and
Engineering of the National Academy of Sciences of Ukraine, Lenin Ave., 47,
Kharkov, 61103, Ukraine}

\date{\today}

\begin{abstract}
The new integrable quantum spin model is proposed. The model has a biaxial
magnetic anisotropy of alternating coupling between spins together with
multiple spin interactions. Our model gives the possibility to exactly find
thermodynamic characteristics of the considered spin chain. The ground state
of the model can reveal spontaneous values of the total magnetic and
antiferromagnetic moments, caused by multiple spin couplings. Also, in the
ground state, depending on the strength of multiple spin couplings, our model
manifests several quantum critical points, some of which are governed
by the external magnetic field.
\end{abstract}

\maketitle

Integrable models of quantum physics of magnetism are,
unfortunately, rare. Such models, however, are very important for
theorists, because they permit to compare the results of standard
for theoretical physics of real systems perturbative approaches
with exact ones. For example, the seminal Ising model served as a
basis for many powerful methods of modern physics, like the
scaling, renormalization group, etc. Quantum integrable models in
one space dimension (1D) are developed relatively more, comparing
to the 2D or 3D ones (in fact there exist only few examples of
quantum integrable models in 2D and 3D), due to the relative
simplicity of their study. On the other hand, according to the
Mermin-Wagner theorem, any nonzero temperature in 1D (and 2D)
destroys long-range magnetic ordering. That theorem reveals, in
fact, the enhancement of quantum and thermal fluctuations in
low-dimensional systems, due to peculiarities in their densities
of states.

On the other hand, the interest in quantum spin systems, where
spin-spin interactions along one or two space directions are much
stronger than couplings along other directions, has considerably
grown during last decade. Such interest to low-dimensional quantum
spin systems is motivated, first of all, by the progress in the
preparation of real substances with well defined one-dimensional
subsystems. On the other hand, modern technologies permit to
compose artificial one-dimensional quantum systems, like quantum
wires and rings, which properties are created to be similar to
theoretically known models. Devices, based on such especially
prepared quantum 1D spin systems are very promising in the modern
nano-technology, in the development of spintronics, or even in the
quantum computation.

1D quantum spin systems often manifest quantum phase transitions, i.e. those,
which take place in the ground state, and which are governed by other than
the temperature parameters, like an external magnetic field, external or
internal (caused by chemical substitutions) pressure, etc.

In the past most of exactly solvable quantum spin models were related to the
class of models with only pair spin-spin interactions between nearest-neighbor
spins. \cite{Zb} Last years more attention of physicists was paid to
theoretical studies of quantum spin models with not only nearest-neighbor
spin-spin interactions, but also with next-nearest neighbor ones, multiple
spin exchange models (e.g., a ring exchange), etc.
\cite{Tsv,BZPS,PZ,MT,obz,TJ,LWC,Z} Such models sometimes appear to be
closer to the real situation in quasi-one-dimensional magnets, comparing to
the ones with only nearest-neighbor couplings between only pairs of spins. For
example, such additional interactions are often present in oxides of
transition metals, where the direct exchange between magnetic ions is
complimented by the superexchange between magnetic ions via nonmagnetic ones.
Terms, involving the product of three and four spin operators or more, were
only recently recognized to be important for the theoretical description of
many physical systems, despite the fact that multiple spin exchange models
were introduced by Thouless already in 1965. \cite{Tho} For example, multiple
spin exchanges were used for the description of the magnetic properties of
solid $^3$He. \cite{He}  Later similar models were used to study some
cuprates \cite{cupr} and spin ladders. \cite{ladd} Often quantum spin models
with antiferromagnetic interactions and multiple spin interactions manifest
the spin frustration, i.e. the lowest in energy state is highly degenerate.
\cite{MBLW} Last but not the least, such models often reveal quantum phase
transitions. For instance, many transition metal compounds, like copper
oxides, are believed to manifest features, characteristic for quantum phase
transitions. However, it is known that for many quantum spin models the
standard quasi-classical theoretical description, based on the quantization
of small deviations of classical vectors of magnetization of magnetic
sublattices, yield incorrect results, especially in the vicinity of quantum
critical points. \cite{KT} This is why, quantum exactly solvable spin models
with multiple spin exchange interactions, even being rather formal, and,
sometimes, non-realistic, are of great importance: They provide the
possibility to check approximate theoretical methods, used for the
description of more realistic physical models of quantum spin systems with
spin frustration.

In this paper we propose a new integrable model of quantum spins
with nearest-neighbor interactions and multiple spin exchange. The
aim of this work is to study a model, that, on the one hand,
contains multiple spin interactions, which usually produce
incommensurate magnetic structures. \cite{obz} Second, the
proposed model consists alternating exchange interactions between
nearest neighbors, which can be the reason for the spin gap for
low-lying excitations. \cite{Zb} Finally, the model has the
biaxial magnetic anisotropy, which is believed to be the key
property of transition metal compounds with strong spin-orbit
coupling. \cite{KH} The 2D counterpart of the model can be related
to the plaquette model of $p_1+ip_2$ superconducting arrays.
\cite{NF} On the other hand, the model is relatively simple,
because the Hamiltonian of the model can be exactly transformed to
the one of the free fermion lattice gas, and, hence, most of
thermodynamic characteristics can be calculated explicitly.

The Hamiltonian of our exactly solvable quantum spin model with alternating
nearest-neighbor couplings and three-spin interactions, which permits exact
solution, has the form:
\begin{eqnarray}
&&{\cal H} = - H\sum_n (\mu_1S^z_{n,1} +\mu_2S^z_{n,2}) -\sum_n
(J_{1x}S^x_{n,1}S^x_{n,2} \nonumber \\
&&+J_{1y}S^y_{n,1}S^y_{n,2}) - \sum_n (J_{2x}S^x_{n,2}S^x_{n+1,1}
+J_{2y}S^y_{n,2}S^y_{n+1,1}) \nonumber \\
&&- J_{13}\sum_n (S^x_{n,1}S^z_{n,2}S^x_{n+1,1} +S^y_{n,1}S^z_{n,2}
S^y_{n+1,1}) \nonumber \\
&&- J_{23}\sum_n (S^x_{n,2}S^z_{n+1,1}S^x_{n+1,2} +S^y_{n,2}S^z_{n+1,1}
S^y_{n+1,2}) \ , \
\label{H1}
\end{eqnarray}
where $S_{n,1,2}^{x,y,z}$ are the operators of the spin-${1\over2}$
projections of the spin in the $n$-th cell, which belongs to the sublattice 1
or 2, $\mu_{1,2}$ are effective magnetons of the sublattices, $H$ is the
external magnetic field, directed along $z$ axis, $J_{1,2,x,y}$ are
the alternating exchange coupling constants between nearest neighbor spins in
the cell and between cells, and $J_{1,23}$ are alternating coupling constants
for three-spin interactions. Notice that the model reveals the biaxial
magnetic anisotropy, i.e. the exchange interactions (in the spin subspace)
along $x$, $y$, and $z$ directions are different. In the case
$J_{1,2x}=J_{1,2y}$, i.e. in the case of only uniaxial magnetic anisotropy,
the model contains, as a special case, the model, studied in
Ref.~\onlinecite{ZS}. On the other hand, the special case $J_{13}=J_{23}=0$
of the model is known for many years. \cite{P} Finally, in the absence of the
magnetic field, $H=0$, and three-spin couplings, $J_{13}=J_{23}=0$ the model
can be related to the so-called 1D quantum compass model (in the special case
$J_{1x}=\alpha$, $J_{1y}=1-\alpha$, $J_{2x}=0$ $J_{2y}=1$). \cite{qc}
Terms, in the Hamiltonian, which describe three-spin couplings, obviously
violate time-reversal and parity symmetries of the system separately, but the
combination of both symmetries is preserved.

\begin{figure}
\begin{center}
\includegraphics[height=7cm,width=0.35\textwidth]{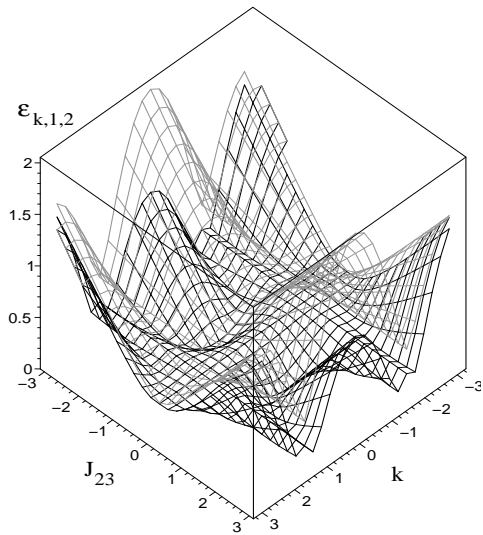}
\end{center}
\caption{Dispersion relations for the exactly solvable quantum compass model
at zero magnetic field for the upper branch (grey) and lower branch (black) of
eigenstates as functions of the parameter of three-spin interactions $J_{23}$
for $J_{13}=J_{23}$.}
\label{qc1}
\end{figure}

\begin{figure}
\begin{center}
\includegraphics[height=7cm,width=0.35\textwidth]{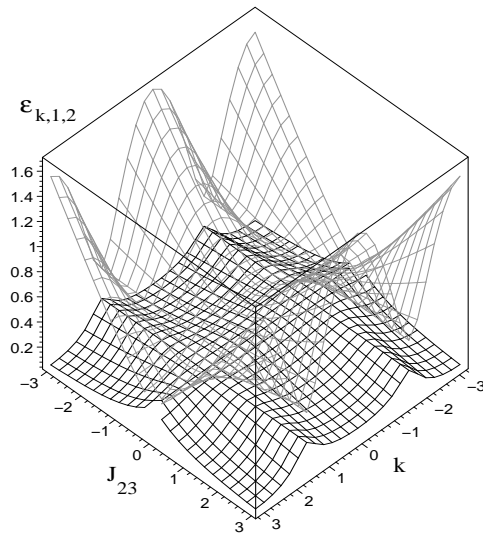}
\end{center}
\caption{The same as in Fig.~1, but for $J_{13}=0$.}
\label{qc2}
\end{figure}

After the Jordan-Wigner transformation \cite{Zb,KT}
\begin{eqnarray}
&&S_{n,1,2}^z = {1\over2}\sigma_{n,1,2} = {1\over2} - a^{\dagger}_{n,1,2}
a_{n,1,2} \ , \nonumber \\
&&S_{n,1}^+ \equiv S_{n,1}^x + iS_{n,1}^y =
\prod_{m<n}\sigma_{m,1}\sigma_{m,2}a_{n,1} \ , \nonumber \\
&&S_{n,1}^- \equiv S_{n,1}^x - iS_{n,1}^y =
a^{\dagger}_{n,1}\prod_{m<n}\sigma_{m,1}\sigma_{m,2} \ , \nonumber \\
&&S_{n,2}^+ =\prod_{m<n}\sigma_{m,1}\sigma_{m,2}\sigma_{n,1}a_{n,2} \ ,
\nonumber \\
&&S_{n,2}^- = a^{\dagger}_{n,2}\prod_{m<n}\sigma_{m,1}\sigma_{m,2}
\sigma_{n,1} \ ,
\label{JW}
\end{eqnarray}
where $a^{\dagger}_{n,1,2}$ and $a_{n,1,2}$ are creation and destruction
operators, which satisfy fermionic anticommutation relations, and, after the
Fourier transform
\begin{equation}
a_{n,1,2} = N^{-1/2}\sum_k a_{k,1,2}\exp (ikn)
\end{equation}
and similar for $a^{\dagger}_{n,1,2}$, where $N$ is the number of cells, the
Hamiltonian Eq.~(\ref{H1}) gets the form
\begin{eqnarray}
&&{\cal H} =  \sum_k \biggl[ \left(\mu_1H -
{J_{13}\over 2}\cos k \right)a^{\dagger}_{k,1}a_{k,1}
\nonumber \\
&&+ \left( \mu_2H -{J_{23}\over 2}\cos k \right)a^{\dagger}_{k,2}a_{k,2}
\nonumber \\
&&- {1\over2}\left(J_1^+ +J_2^+e^{-ik} \right)a^{\dagger}_{k,1}a_{k,2}
\nonumber \\
&&- {1\over2}\left(J_1^+ +J_2^+e^{ik} \right)a^{\dagger}_{k,2}a_{k,1}
\nonumber \\
&&- {1\over2}\left(J_1^- -J_2^-e^{-ik} \right)a^{\dagger}_{k,1}
a^{\dagger}_{k,2}  \nonumber \\
&&- {1\over2}\left(J_1^- -J_2^-e^{ik} \right)a_{k,2}a_{k,1} \biggr]
- {\mu_1+\mu_2\over2}NH \ ,
\label{H2}
\end{eqnarray}
where $J_{1,2}^{\pm} = (1/2)(J_{1,2x} \pm J_{1,2y})$. With the help of a
unitary transformation this Hamiltonian can be diagonalized
\begin{equation}
{\cal H} = \sum_k\sum_{j=1}^2 \varepsilon_{k,j}\left(b^{\dagger}_{k,j}
b_{k,j} - {1\over 2} \right) \ ,
\label{H3}
\end{equation}
where
\begin{equation}
\varepsilon_{k,1,2}^2 = F_k \pm \sqrt{F^2_k -G_k} \ ,
\label{disp}
\end{equation}
and
\begin{eqnarray}
&&F_k ={1\over8}\biggl(J_{1x}^2+J_{1y}^2+J_{2x}^2+J_{2y}^2
\nonumber \\
&&+2(J_{1x}J_{2y}+J_{1y}J_{2x})\cos k + (J_{13}^2+J_{23}^2)\cos^2k
\nonumber \\
&&+4(\mu_1^2+\mu_2^2)H^2 -4(\mu_1J_{13}+\mu_2J_{23})H\cos k \biggr) \ ,
\nonumber \\
&&G_k = \biggl( \mu_1\mu_2H^2 -{1\over 2}(\mu_1J_{23} +\mu_2J_{13})H\cos k
\nonumber \\
&&+{1\over 4}[J_{13}J_{23}\cos^2k -J_{1x}J_{1y} -J_{2x}J_{2y}
\nonumber \\
&&-(J_{1x}J_{2x}+J_{1y}J_{2y})\cos k] \biggr)^2
\nonumber \\
&&+ {1\over 16}(J_{1x}J_{2x}-J_{1y}J_{2y})^2\sin^2k \ .
\label{FG}
\end{eqnarray}
Using the standard particle-hole transformation one can get only positive
eigenvalues of the Hamiltonian (\ref{disp}). One can check that in the case
$J_{1,23}=0$ the spectrum coincides with the one of Ref.~\onlinecite{P}, while
for $J_{1,2x}=J_{1,2y}$ it reproduces the spectrum from Ref.~\onlinecite{ZS}.
The energies of eigenstates of the first branch are non-negative for all
parameters of the model. The energies of eigenstates, belonging to the second
branch can be equal to zero, depending on the values of the parameters of the
model. Figs.~\ref{qc1}-\ref{qc2} represent the zero field dispersion relations
for both branches as functions of three-spin interactions for the homogeneous
three-spin couplings and for the alternating three-spin couplings,
respectively, for the quantum compass model with $\alpha =0.4$

Despite some parameter-dependent features, the behavior of our
model is similar for the quantum compass case with very strong
biaxial magnetic anisotropy and the case with small biaxial
anisotropy.

It is simple to obtain thermodynamic characteristics of our model at nonzero
temperatures. The free energy of the quantum spin chain is equal to
\begin{equation}
F = -T \sum_{k}\sum_{j=1}^2 \ln \left(2\cosh {\varepsilon_{k,j}\over2T}
\right) \ .
\label{F}
\end{equation}
Obviously, the $z$-projection of the average magnetization of the system is
\begin{equation}
M^z = {1\over2}\sum_{k}\sum_{j=1}^2 {\partial \varepsilon_{k,j}\over \partial
H}\tanh \left( {\varepsilon_{k,j}\over 2T}\right) \ .
\end{equation}
From this formula it is easy to show that $M^z$ is zero for $H = 0$ for any
nonzero temperature, in accordance with the Mermin-Wagner theorem. The low
temperature behavior of the magnetic susceptibility,
\begin{eqnarray}
&&\chi = {1\over2}\sum_{k}\sum_{j=1}^2 \biggl[{\partial^2 \varepsilon_{k,j}
\over \partial H^2}\tanh \left( {\varepsilon_{k,j}\over 2T}\right)
\nonumber \\
&&+ \left({\partial \varepsilon_{k,j}\over \partial
H}\right)^2 \left[2T\cosh \left({\varepsilon_{k,j}\over
2T}\right)\right]^{-2}\biggr]^{-1} \ ,
\end{eqnarray}
and the specific heat,
\begin{equation}
c = \sum_k \sum_{j=1}^2 {\varepsilon_{k,j}^2\over 4T^2\cosh^2
(\varepsilon_{k,j}/2T)}
\end{equation}
depend on the values of coupling constants $J_{1,2}^{x,y}$,
$J_{13,23}$, the effective magnetons, $\mu_{1,2}$, and the value
of the external magnetic field $H$, see below. One can check that
there is no ordering and, therefore, none of thermodynamic
characteristics of the considered system has peculiarities at any
nonzero temperature. On the other hand, as it will be shown below,
in the ground state spontaneous magnetic ordering can take place.
In the cases, where elementary excitations of the model are
gapped, the low-temperature magnetic susceptibility and the
specific heat reveal exponential in $T$ dependencies in the
absence of spontaneous magnetization. If the model reveals the
spontaneous magnetic moment, the magnetic susceptibility at low
temperatures is divergent. On the other hand, for gapless
situation of low-energy states of the model, the magnetic
susceptibility is finite at low temperatures for the absence of
spontaneous magnetic ordering at $T=0$, while the specific heat is
linear in $T$. At the critical lines of quantum phase transitions
(see below) our model manifests either square root, or logarithmic
in $T$ and magnetic field behaviors of the specific heat and the
magnetic susceptibility. In the case, where interaction constants
are very different from each other (or, to be more precise, when
two branches of eigenstates are characterized by very different
energy scales), the specific heat and the magnetic susceptibility
can reveal two-maxima temperature dependencies.

The most important properties of the one-dimensional spin system are
manifested in the ground state. The ground state energy of our model can be
written as
\begin{equation}
E_0=-{1\over \sqrt{2}} \sum_{k}\sqrt{F_k +\sqrt{G_k}} \ .
\label{gstate}
\end{equation}
Then the $z$-projections of each total spin moment of a cell in the ground
state can be written as:
\begin{equation}
S^z_{1,2} \equiv {\partial E_0\over \partial \mu_{1,2}H} =
{1\over 4\sqrt{2}}\sum_k {2\sqrt{G_k} x_{1,2,k} +y_{1,2,k} \over
\sqrt{G_k} \sqrt{F_k+\sqrt{G_k}}} \ ,
\label{S12}
\end{equation}
where
\begin{eqnarray}
&&x_{1,2,k} = \mu_{1,2}H -{1\over 2}J_{1,23}\cos k \ ,
\nonumber \\
&&y_{1,2,k}= \biggl( 4\mu_1\mu_2H^2 -(\mu_1J_{23} +\mu_2J_{13})H\cos k
\nonumber \\
&&+{1\over 2}[J_{13}J_{23}\cos^2k -J_{1x}J_{1y} -J_{2x}J_{2y}
\nonumber \\
&&-(J_{1x}J_{2x}+J_{1y}J_{2y})\cos k] \biggr)
\nonumber \\
&&\times (\mu_{2,1}H-{1\over2}J_{2,13}\cos k) \ .
\label{xy12}
\end{eqnarray}
The sum of the $z$-projections of spin moments can be considered
as the ground state vector of magnetism, or magnetization of the
model, $M^z=\mu_1S_1^z +\mu_2S_2^z$, while the difference
describes the vector of antiferromagnetism, \cite{P} $L^z =
\mu_1S^z -\mu_2S_2^z$, or the staggered magnetization of the
model. From these expressions one immediately sees that in the
ground state the model can have nonzero spontaneous magnetic and
antiferromagnetic moments (i.e. magnetic ordering for $H=0$),
caused by nonzero three-spin interactions. We would like to turn
attention that the signs of $J_{1,23}$ do matter. Namely,
depending on their signs, the spontaneous magnetization of the
model in the ground state can be positive or negative, with
respect to the direction of the magnetic field. It is different
from the behaviors of other known exactly solvable spin chain
models. The reason for the onset of the spontaneous magnetic and
antiferromagnetic moments in the ground state of our model is
related to the violation of the time-reversal symmetry by
three-spin coupling terms.

It is interesting to notice that the equality $G_k=0$ means that
$\varepsilon_{k,2} =0$. As it is shown below, namely the condition $G_k=0$
determines the features in the behavior of all ground state characteristics
of the spin chain. One can see, that $G_k=0$ either at $\sin k =0$ (i.e. for
$k=0, \pi$), or, for any $k$, if $J_{1x}J_{2x}=J_{1y}J_{2y}$ (it turns out
that this condition does not depend on the magnetic field and on the values
of three-spin couplings).

\begin{figure}
\begin{center}
\includegraphics[height=7cm,width=0.35\textwidth]{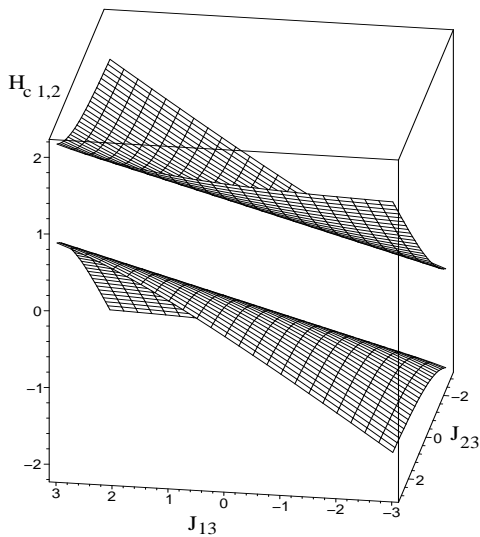}
\end{center}
\caption{Critical values of the magnetic fields $H_{c1,2}$ for the exactly
solvable spin model as functions of parameters of three-spin interactions
$J_{13}$ and $J_{23}$. We used $\mu_1=1.01$, $\mu_2=0.99$, $J_{1x}=1$,
$J_{1y}=1.5$, $J_{2x}=2$, and $J_{2y}=0.9$.}
\label{H12}
\end{figure}

\begin{figure}
\begin{center}
\includegraphics[height=7cm,width=0.35\textwidth]{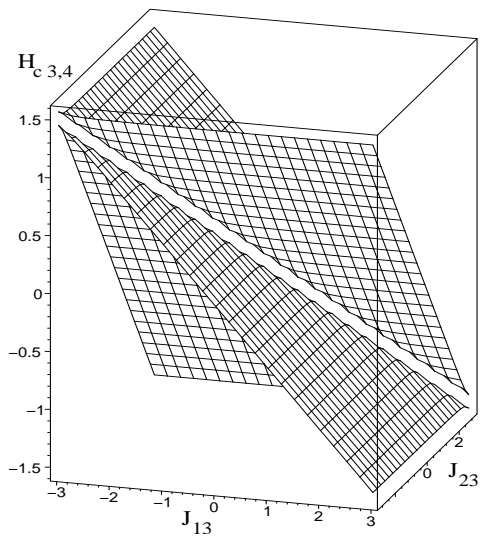}
\end{center}
\caption{Critical values of the magnetic fields $H_{c3,4}$ for the exactly
solvable spin model as functions of parameters of three-spin interactions
$J_{13}$ and $J_{23}$. The set of parameters is the same as in Fig.~3.}
\label{H34}
\end{figure}

Let us consider first the case with $J_{1x}J_{2x}\ne J_{1y}J_{2y}$. Notice
that the limiting case of the quantum compass model belongs to the situation.
The first branch of eigenstates is ever positive, but the second one can reach
zero only for two values of the quasimomenta ($k=0, \pi$). Then it is simple
to show that the ground state magnetisation is a continuous function of the
external magnetic field, except of at $H=0$ for $\mu_1J_{23} \ne \mu_2J_{13}$
and $\mu_1J_{13} \ne \mu_2J_{23}$, see Eqs.~(\ref{S12}-\ref{xy12}). For the
latter the spontaneous magnetization appears, and, therefore, the ground state
magnetic susceptibility is divergent there at $H=0$. The magnetic
susceptibility for nonzero values of $H$ can have peculiarities, proportional
to $\ln |H-H_{c,i}|$ ($i=1,2,3,4$), if the external magnetic field becomes
equal to one of following four values
\begin{eqnarray}
&&H_{c,1,2} = {1\over 4\mu_1\mu_2}\biggl( (\mu_1J_{23}+\mu_2J_{13})
\pm \bigg[(\mu_1J_{23} -\mu_2J_{13})^2
\nonumber \\
&&+\mu_1\mu_2(J_{1x} +J_{2y})(J_{1y} +J_{2x})\biggr]^{1/2} \ ,
\nonumber \\
&&H_{c,3,4} = {1\over 4\mu_1\mu_2}\biggl( -(\mu_1J_{23}+\mu_2J_{13})
\pm \bigg[(\mu_1J_{23}
\nonumber \\
&&-\mu_2J_{13})^2 +\mu_1\mu_2(J_{1x} -J_{2y})(J_{1y}
-J_{2x})\biggr]^{1/2} \biggr) \ .
\label{Hc1}
\end{eqnarray}
at which second order quantum phase transitions can take place, see
Figs.~\ref{H12}-\ref{H34}.

Such quantum critical points exist, naturally, only if those critical values
of the field are real and non-negative. They are real if
\begin{equation}
|\mu_1\mu_2(J_{1x} \pm J_{2y})(J_{1y} \pm J_{2x})| \le
(\mu_1J_{23} -\mu_2J_{13})^2 \ ,
\label{condr}
\end{equation}
and the first two critical values are non-negative for $\mu_1\mu_2 > 0$, if
\begin{eqnarray}
&&(\mu_1J_{23} +\mu_2J_{13}) \ge \biggl[ (\mu_1J_{23} -\mu_2J_{13})^2
\nonumber \\
&&+\mu_1\mu_2(J_{1x} +J_{2y})(J_{1y} +J_{2x})\biggr]^{1/2} > 0 \ ,
\label{cond1}
\end{eqnarray}
or the second two critical values are non-negative, if
\begin{eqnarray}
&&-(\mu_1J_{23} +\mu_2J_{13}) \ge \biggl[ (\mu_1J_{23} -\mu_2J_{13})^2
\nonumber \\
&&+\mu_1\mu_2(J_{1x} -J_{2y})(J_{1y} -J_{2x})\biggr]^{1/2} > 0 \ .
\label{cond2}
\end{eqnarray}
For $\mu_1\mu_2 < 0$ the non-negativity conditions are reversed.
If the reality conditions are not satisfied, no quantum phase transitions,
governed by the external magnetic field, take place in the system. If one of
them is satisfied, and the other isn't, then only up to two quantum phase
transitions can happen. If the conditions Eqs.~(\ref{cond1}), or
(\ref{cond2}), are not satisfied, then only one or two quantum phase
transitions, governed by the field, take place. If one of the effective
magnetons is zero (i.e. one of the ions, which form elementary cell, is
non-magnetic), but three-spin interaction constants are not, then the quantum
phase transitions take place at the values of the magnetic field
\begin{equation}
H_{c5,6} =  \pm {J_{13}J_{23} + (J_{1x} \pm J_{2y})
(J_{1y} \pm J_{2x})\over 2\mu_{1,2}J_{2,13}} \ ,
\label{Hc2}
\end{equation}
at which the magnetic susceptibility has logarithmic
singularities. Naturally, only positive values of $H_{c5,6}$
matter. Finally, if both of effective magnetons are zero, then,
obviously, there are no quantum phase transitions, governed by the
magnetic field. It turns out that at nonzero temperatures
thermodynamic characteristics of the model reveal logarithmic in
$T$ features at the critical values of the magnetic field.

Consider now the situation, in which $J_{1x}J_{2x}= J_{1y}J_{2y}$. The
energies of the eigenstates in this case can be written as
\begin{eqnarray}
&&\varepsilon_{k,1,2} = \biggl[ \bigl({1\over4} \left[(\mu_1+\mu_2)H -
{1\over 2} (J_{13}+J_{23})\cos k \right]^2
\nonumber \\
&&+ |B_k|^2 \bigr)^{1/2} \pm \bigl( {1\over4}
\left[(\mu_1-\mu_2)H -{1\over 2}(J_{13}-J_{23})\cos k \right]^2
\nonumber \\
&&+ |A_k|^2 ]\bigr)^{1/2} \biggr]^{1/2}
\ ,
\label{disp2}
\end{eqnarray}
where
\begin{eqnarray}
&&A_k ={1\over 2} [J_1^+ +J_2^+ \exp (-ik)] \ ,
\nonumber \\
&&B_k ={1\over 2} [J_1^- -J_2^-\exp (-ik)] \ .
\label{AB}
\end{eqnarray}
It is obvious that $\varepsilon_{k,1}$ is positive for any parameters of the
model. On the other hand, for the lower branch for some ranges of the
quasimomentum $k$ and external magnetic field $H$, the first term under the
square root sign in Eq.~(\ref{disp2}) can be smaller than the second one. It
implies that eigenstates for lower branch can exist only for some ranges of
$k$, depending on the value of the external field $H$. The analysis of this
situation is similar to the above (except the fact that one has to take into
account nonzero Fermi seas, i.e. totally filled states with negative energies
for some ranges of $k$ depending on the value of the external field; the
critical value of $k$ is determined from the condition
$\varepsilon_{2,k_c}=0$). One can see that there exist four critical values
of the magnetic field, at which quantum phase transitions can take place, see
Eqs.(\ref{Hc1})-(\ref{Hc2}). The difference, comparing to the case with
$J_{1x}J_{2x}\ne J_{1y}J_{2y}$, is in the more strong features of the
magnetic susceptibility at critical fields $\sim 1/\sqrt{|H-H_{c,i}|}$ in the
ground state, and, therefore, in square root peculiarities in $T$ of
thermodynamic characteristics of the model, like the magnetic susceptibility
and specific heat, at critical values of the magnetic field.

Let us consider the homogeneous limiting case of our model $J_{1x}=J_{2x}=Jx$,
$J_{1y}=J_{2y}=J_y$, $J_{13}=J_{23}=J_3$, $\mu_1=\mu_2=\mu$. In this case the
Hamiltonian can be written as
\begin{equation}
{\cal H} = \sum_k\varepsilon_{k}\left(b^{\dagger}_{k}
b_{k} - {1\over 2} \right) \ ,
\label{H3h}
\end{equation}
where
\begin{eqnarray}
&&\varepsilon_{k}^2 = \left[ \mu H -{1\over2}(J_3\cos (2k) + (J_x+J_y)\cos
(k) \right]^2 \nonumber \\
&&+ {1\over 4}(J_x-J_y)^2\sin^2(k) \ .
\label{disph}
\end{eqnarray}
One can see that the energy (\ref{disph}) is non-negative. It can be equal to
zero only for $J_x=J_y$, or, if $J_x\ne J_y$ for $k =0, \pi$. In the later
case there are two critical values of the magnetic field, at which quantum
phase transitions can take place
\begin{equation}
H_{c,1,2}^h = (2\mu)^{-1}[J_3 \pm (J_x+J_y)] \ .
\label{Hch}
\end{equation}
Obviously, quantum phase transitions take place if values of the critical
field are non-negative, i.e. they take place for $\mu >0$, if
\begin{equation}
J_3 \pm (J_x+J_y) \ge 0
\label{condh}
\end{equation}
Hence, $J_3 =\pm (J_x+J_y)$ are the conditions of the quantum phase
transition, governed by the three-spin coupling. The ground state magnetic
susceptibility has logarithmic features $\sim \ln |h-H^h_{c1,2}|$ at
critical values of the field.

\begin{figure}
\begin{center}
\includegraphics[height=7cm,width=0.38\textwidth]{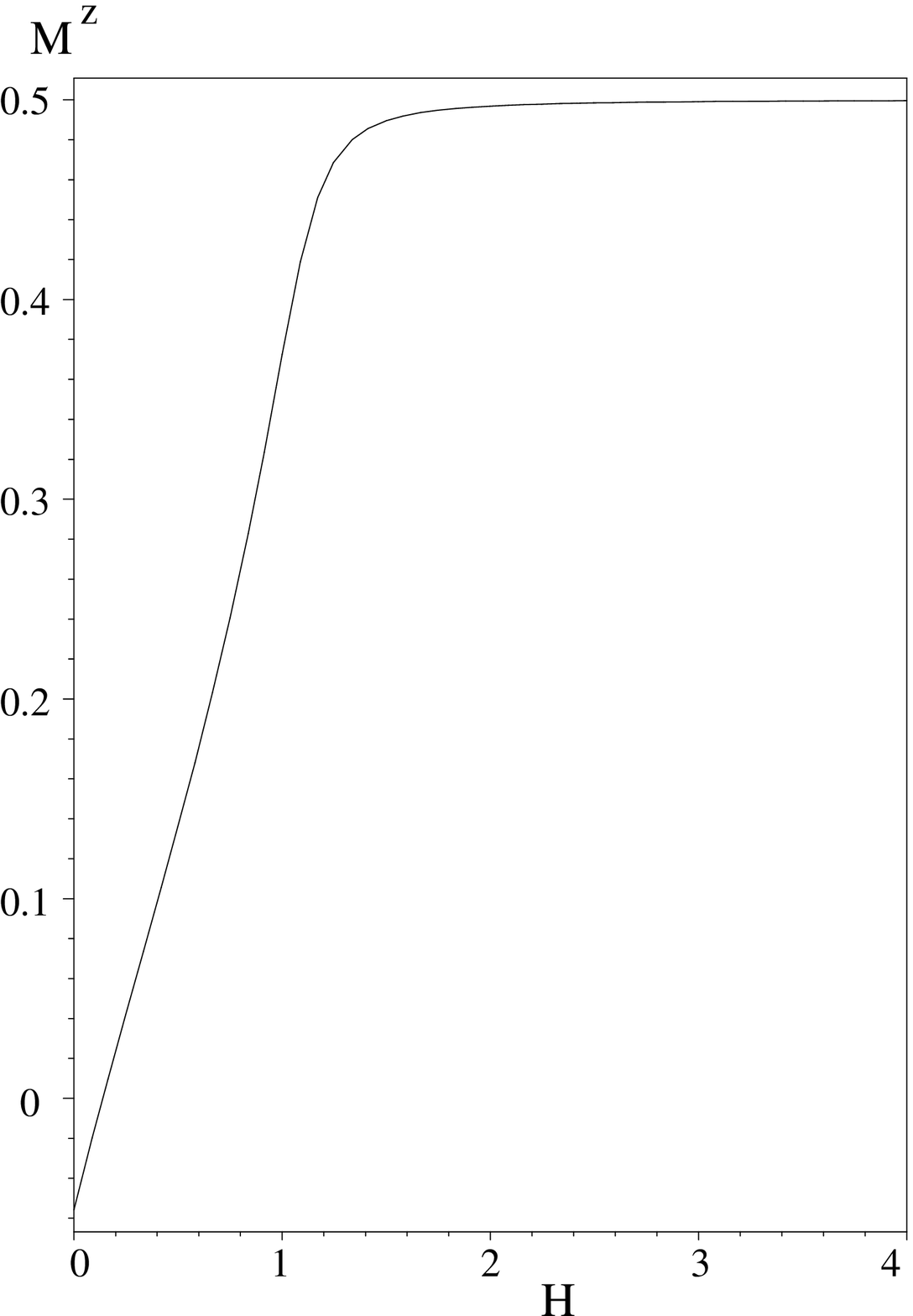}
\end{center}
\caption{The ground state dependencies of the magnetization as a
function of the magnetic field for the homogeneous limit of the
exactly solvable model for $J_x=1$, $J_y=0.6$ for $J_3=-2$.}
\label{mz10}
\end{figure}

\begin{figure}
\begin{center}
\includegraphics[height=7cm,width=0.38\textwidth]{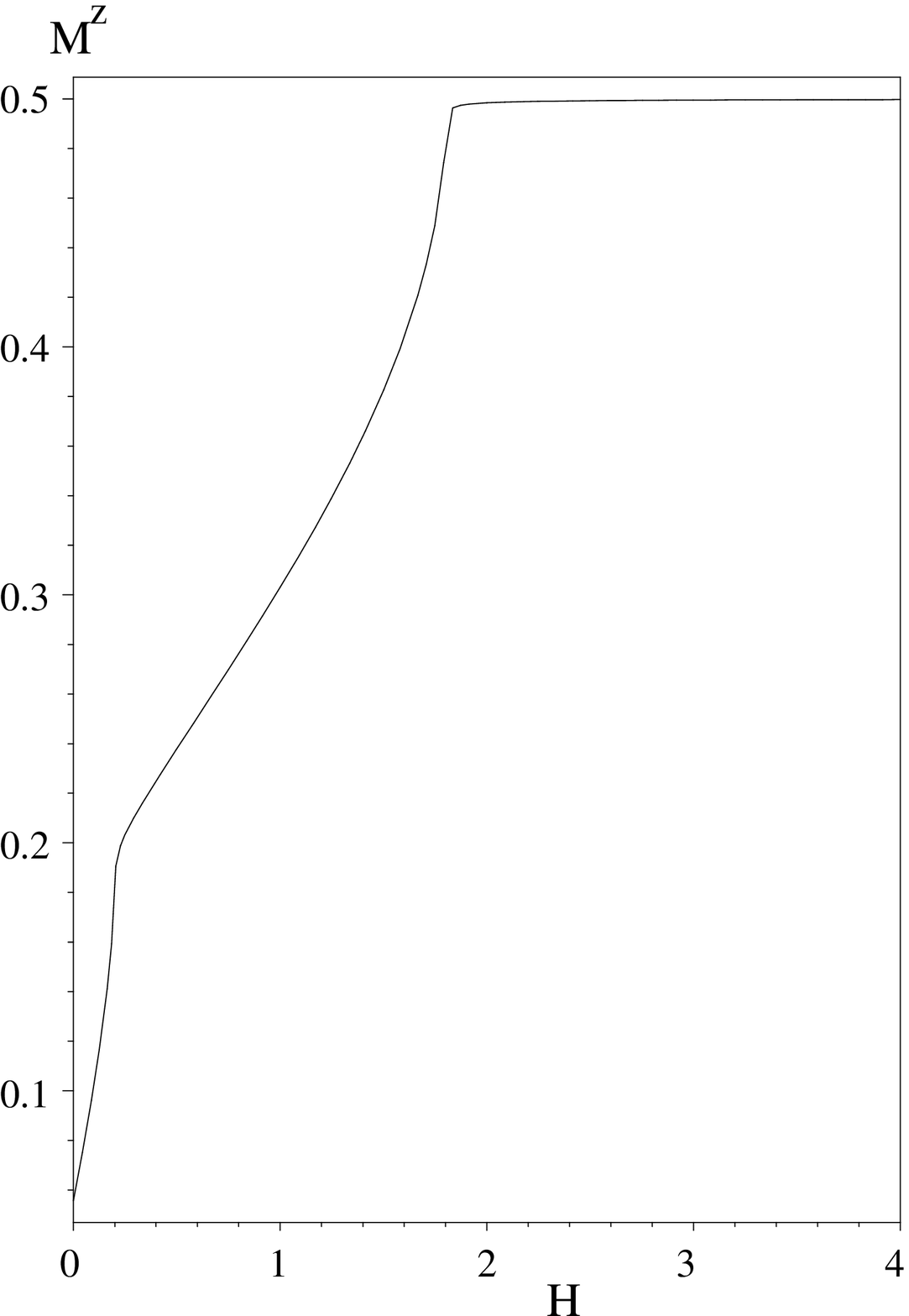}
\end{center}
\caption{The ground state dependencies of the magnetization as a
function of the magnetic field for the homogeneous limit of the
exactly solvable model for $J_x=1$, $J_y=0.6$ for $J_3=2$.}
\label{mz12}
\end{figure}

Figs.~\ref{mz10} and \ref{mz12} show the ground state behavior of the
magnetization of our model for the homogeneous case as a function of the
magnetic field. Fig.~\ref{mz10} presents the behavior for $J_3 < -(J_x+J_y)$,
and Fig.~\ref{mz12} demonstrates the magnetic field behavior for the region
$J_3 > (J_x+J_y)$. One can see that for both regions there is a spontaneous
magnetization, but its sign (with respect to the direction of the field)
depends on the sign of three-spin interactions. Also, there are two quantum
phase transitions for $J_3 > (J_x+J_y)$, while for $J_3 < -(J_x+J_y)$ the
ground state magnetization is a smooth function of $H$.

\begin{figure}
\begin{center}
\includegraphics[height=7cm,width=0.38\textwidth]{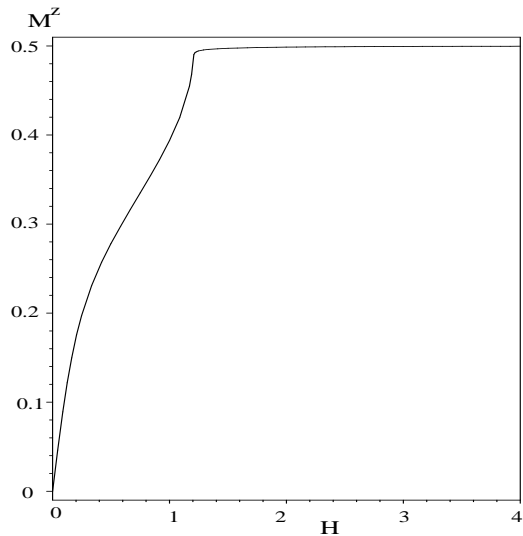}
\end{center}
\caption{The ground state dependencies of the magnetization as a
function of the magnetic field for the homogeneous limit of the
exactly solvable model for $J_x=1$, $J_y=0.6$ for $J_3=0.2$.}
\label{mz2a}
\end{figure}

\begin{figure}
\begin{center}
\includegraphics[height=7cm,width=0.38\textwidth]{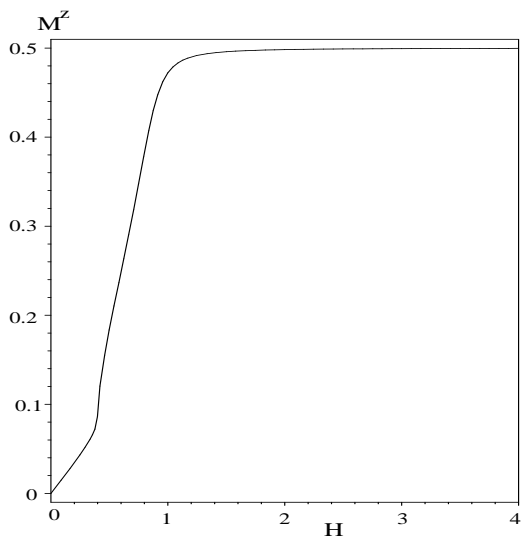}
\end{center}
\caption{The ground state dependencies of the magnetization as a
function of the magnetic field for the homogeneous limit of the
exactly solvable model for $J_x=1$, $J_y=0.6$ for $J_3=-0.8$.}
\label{mz2d}
\end{figure}

Figs.~\ref{mz2a} and \ref{mz2d} present the magnetic field behavior of the
magnetization for $-(J_x+J_y) < J_3 < (J_x +J_y)$ for positive and negative
values of $J_3$, respectively. One can see that in this region there is no
spontaneous magnetization, and only one second order quantum phase transition
takes place.

On the other hand, if $J_x=J_y$, the eigenstates of the Hamiltonian can be
negative for some ranges of $k$ depending on the value of the magnetic field.
Negative energies imply the nonzero Fermi sea, where eigenstates with
negative energies are totally filled, and the ones with positive energies are
empty. In that case quantum phase transitions yield square root singularities
$\sim 1/\sqrt{|H-H_{c,1,2}^h|}$ of the magnetic susceptibility, cf.
Refs.~\onlinecite{TJ,LWC}.

It is important to point out that the quasiclassical description
of our model (when one replaces spin operators by classical
vectors, and quantizing small deviations from the classical
minimal energy state) does not reproduce exact results for the
inhomogeneous (dimerized) situation. Namely, in the classical
description of the model without biaxial anisotropy one of the
branches of eigenstates is obviously gapless, unlike the exact
result.

In conclusion, motivated by recent experiments on quasi-1D quantum
spin systems and recent theories of quantum compass model, we
proposed the integrable model, in which exchange interactions
between neighboring spins is accompanied by the multiple spin
exchange with the biaxial magnetic anisotropy. The model is simple
(due to the exact mapping to the problem of the lattice free
fermion gas), and, therefore, permits to obtain exactly
thermodynamic characteristics of the considered quantum spin
chain. The most important behavior of the model is in the ground
state. Our model manifests a ferrimagnetic-like ordering in the
ground state. Depending on the signs of the parameters of
three-spin couplings, the spontaneous magnetic moment of the
system in the ground state can be positive or negative (with
respect to the direction of the magnetic field). The system can
undergo several second order quantum phase transitions, governed
by the external magnetic field and the three-spin couplings
strengths (the later can be caused by an external or internal
pressure). Despite some artificial structure of our model, we
expect that more realistic quantum biaxial spin systems with
multiple exchange interactions and the alternation of the exchange
between nearest neighbor spins, will show similar to our simple
model behavior, i.e. our exact solution has generic features for
this class of quantum systems.

Partial support from the Institute of Chemistry of V.N.~Karasin
Kharkov National University is acknowledged.

\end{document}